 \documentclass[twocolumn,showpacs,preprintnumbers,amsmath
  ,amssymb,superscriptaddress]{revtex4}
\usepackage{graphicx}
\usepackage{dcolumn}
\usepackage{bm}
\begin{document}


\title{Ground-state electric quadrupole moment of $^{31}$Al}
\newcommand{\TITech}{Department of Physics, Tokyo Institute of Technology, 
  2-12-1 Oh-okayama, Meguro-ku, Tokyo 152-8551, Japan}
\newcommand{\RIKEN}{RIKEN Nishina Center,
  2-1 Hirosawa, Wako, Saitama 351-0198, Japan}
\newcommand{\Rikkyo}{Department of Physics, Rikkyo University,
  3-34-1 Nishi-Ikebukuro, Toshima-ku, Tokyo 171-8501, Japan}
\newcommand{\JAERI}{Japan Atomic Energy Research Institute,
	Tokai-mura, Ibaraki 319-1195, Japan.}
\newcommand{\Cyric}{Cyclotron and Radioisotope Center, Tohoku University,
6-3 Aoba, Aramaki, Aoba-ku, Sendai, Miyagi, 980-8578, JAPAN}
\newcommand{\SPringEight}{Japan Synchrotron Radiation Research
  Institute (JASRI/SPring-8), 
	1-1-1 Kouto, Sayo, Hyogo 679-5198, Japan}
\author{D.~Nagae}    \email{nagae.daisuke@jaea.go.jp}
                     \altaffiliation[Present Address: ]{\JAERI}
                        \affiliation{\TITech}
\author{H.~Ueno}        \affiliation{\RIKEN}
\author{D.~Kameda}      \affiliation{\RIKEN}
\author{M.~Takemura}    \affiliation{\TITech}
\author{K.~Asahi}       \affiliation{\TITech}
\author{K.~Takase}      \affiliation{\TITech}
\author{A.~Yoshimi}     \affiliation{\RIKEN}
\author{T.~Sugimoto}    \altaffiliation[Present Address: ]{\SPringEight}
                        \affiliation{\RIKEN}
\author{K.~Shimada}     \altaffiliation[Present Address: ]{\Cyric}
                        \affiliation{\TITech}
\author{T.~Nagatomo}    \affiliation{\RIKEN}
\author{M.~Uchida}      \affiliation{\TITech}
\author{T.~Arai}        \affiliation{\TITech}
\author{T.~Inoue}       \affiliation{\TITech}
\author{S.~Kagami}       \affiliation{\TITech}
\author{N.~Hatakeyama}       \affiliation{\TITech}
\author{H.~Kawamura}    \altaffiliation[Also at ]{\RIKEN}
                        \affiliation{\Rikkyo}
\author{K.~Narita}      \affiliation{\Rikkyo}
\author{J.~Murata}      \affiliation{\Rikkyo}
\date{\today}
\begin{abstract}
Ground-state electric quadrupole moment of $^{31}{\rm Al}$ 
($I^{\pi}$ =5/2$^{+}$, $T_{1/2}=644(25)$~ms) has been measured by
means of the $\beta$-NMR spectroscopy using a spin-polarized $^{31}$Al
beam produced in the projectile fragmentation reaction. The obtained
$Q$ moment, $| Q_{\rm exp}(^{31}{\rm Al}) |$ = 112(32)~$e$mb, are in
agreement with conventional shell model calculations within the 
{\it sd} valence space. Previous result on the magnetic moment also
supports the validity of the {\it sd} model in this isotope, and thus
it is concluded that $^{31}{\rm Al}$ is located outside of the 
{\it island of inversion}.  
\end{abstract}
\maketitle
The ground states of Ne, Na, and Mg isotopes with neutron numbers
around the magic number $N$ = 20 have been known to show anomalously
tight bindings since 1970's~\cite{THIBAULT,DETRAZ83}. Later,
spectroscopic studies have revealed that the first exited $2^{+}$
levels are lowered~\cite{DETRAZ79,GUILLEMAUD} and their $B$(E2) values
are enhanced~\cite{MOTOBAYASHI} sizably in these isotopes, and the
possibility of deformation has been proposed.  Theoretical
analyses~\cite{II_WARBURTON} discussed the importance of 
2{\it p}-2{\it h} excitations from the {\it sd} shell to the upper
{\it pf} shell, and concluded it plausible that an inversion of
amplitudes between the {\it sd} normal and {\it pf} intruder
configurations would lead to deformation of the ground states. The
region of nuclei where such a phenomenon occurs is called the 
{\it island of inversion}. In elucidating the underlying mechanism for
the inversion, the measurements of the electromagnetic moments have
played an important role. For example in a series of neutron-rich Na
isotopes, it has been found that, once entering the 
{\it island of inversion}, the ground-state magnetic dipole moment
$\mu$ and electric quadrupole moment $Q$~\cite{HU78_gNa,KE00_Q2629Na}
show clear deviations from the conventional shell-model
predictions~\cite{W}, indicating that $\mu$ and $Q$ are sensitive to
changes in the nuclear configuration~\cite{UT03_Na}. Also in the
recent study of Mg isotopes, anomalous ground-state properties have
been revealed through the $\mu$-moment
measurements~\cite{NE05_g31Mg,YO07_g33Mg}.

In the present work, the ground-state $Q$ moment of $^{31}{\rm Al}$
($I^{\pi}$ =5/2$^{+}$, $T_{1/2}=644(25)$~ms) has been measured by
means of the $\beta$-ray detected nuclear magnetic resonance
($\beta$-NMR) spectroscopy~\cite{SU66_BNMR} applied on a projectile
fragment $^{31}{\rm Al}$ implanted in an 
$\alpha{\mathchar`-}{\rm Al_{2}O_{3}}$ (corundum) single crystal in
which a non-zero electric field gradient acts. A spin-polarized
radioactive-isotope beam (RIB) of $^{31}{\rm Al}$ was obtained from
the projectile fragmentation reaction~\cite{POL}. 
Since the neutron-rich aluminum isotopes are located in the
neighborhood of the {\it island of inversion}, their electromagnetic
moments would signify the possible onset of evolution in the nuclear
structure that ultimately leads to the {\it inversion} phenomenon. So
far, the $\mu$ moments of $^{31-34}$Al~\cite{BO02,HI06} and
$^{30,\ 32}$Al~\cite{UENO_g3032Al} have been reported. The obtained
values of $\mu$ for $^{30-32}$Al seem to stay within the conventional
{\it sd}-model predictions~\cite{W,OXBASH}, indicating that their
structures are suitably described within the normal {\it sd} model
space. Those of $^{33,\ 34}$Al having neutron numbers $N$ = 20 and 21,
on the other hand, seem to indicate deviations from the 0$\hbar\omega$
shell-model predictions~\cite{HI06,HI08}. Since the 
{\it island of inversion} is considered to involve the nuclear
deformation, the $Q$ moment would be a more suitable probe. It has
been found in a recent measurement that $^{32}$Al has a very small $Q$
moment $| Q_{\rm exp}(^{32}{\rm Al}) |$ =
24(2)~$e$mb~\cite{KAMEDA_Q32Al} characteristic of a simple 
($\pi{d_{5/2}}^{-1}\otimes\nu{d_{3/2}}^{-1}$)$^{J=1}$ configuration,
indicating a spherical shape. The $Q$ moment of $^{31}{\rm Al}$ is
important in elucidating how the nuclear shape evolves along the
aluminum isotopes toward and beyond the $^{32}$Al nuclide.

\begin{figure}[tbhp]
\centering
\includegraphics[clip, width=0.4\textwidth]{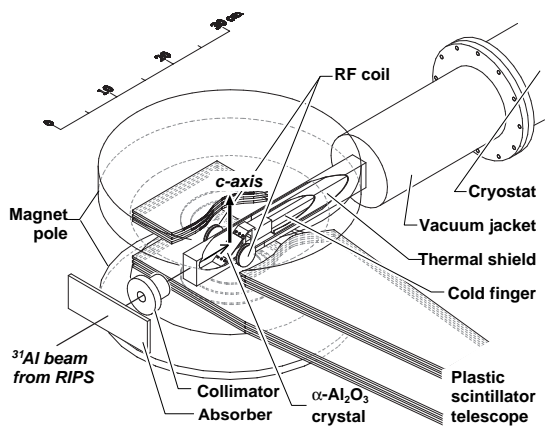}
\caption{Schematic layout of the $\beta$-NMR apparatus.} 
\label{FIG:NQR}
\end{figure}

The experiment was carried out using the RIKEN projectile fragment
separator RIPS~\cite{RIPS}. The arrangement of RIPS for producing the
spin-polarized RIB is essentially the same as that described in
Ref.~\cite{KAMEDA_Q32Al}. A beam of $^{31}{\rm Al}$ was obtained from
the fragmentation of $^{40}$Ar projectiles at $E$ = 95$A$~MeV on a
$0.37$~g/cm$^{2}$-thick $^{93}$Nb target. It has been revealed that a
spin-polarized RIB is obtained in the projectile fragmentation
reaction simply by selecting the angle and momentum of the outgoing
fragments~\cite{POL}. Thus, $^{31}{\rm Al}$ fragments emitted at
angles $\theta_{\rm Lab.}$ = (1.3 -- 5.7)$^{\circ}$ from the primary
beam direction were accepted by RIPS using a beam swinger installed
upstream of the target. Also, a range of momenta $p$ = 
(1.01 -- 1.07)$p_{0}$ was selected with a slit placed at the
momentum-dispersive intermediate focal plane. Here $p_{0}$ =
12.2~GeV/$c$ is the fragment momentum corresponding to the projectile
velocity. The isotope separation was provided by combined analyses of
the magnetic rigidity and momentum loss in the wedge-shaped
degrader~\cite{RIPS}. Then, the spin-polarized $^{31}{\rm Al}$ were
transported to a $\beta$-NMR apparatus located at the final focus of
RIPS, and were implanted in a stopper of 
$\alpha{\mathchar`-}{\rm Al_{2}O_{3}}$ single crystal of hexagonal
structure.  A static magnetic field $B_{0}$ = 501.768(3)~mT was
applied to the stopper. The layout of the $\beta$-NMR apparatus is
shown in Fig.~\ref{FIG:NQR}.  
The $\alpha{\mathchar`-}{\rm Al_{2}O_{3}}$ crystal was cut into a
18~mm$\times$32~mm$\times$0.6~mm slab, and was mounted in a stopper
chamber so that the {\it c}-axis was oriented parallel to the $B_0$
field. The stopper was kept in vacuum and cooled to a temperature $T$
= (70 $\sim$ 100)~K to suppress the spin-lattice relaxation of
$^{31}{\rm Al}$ during the $\beta$ decay. 

The $Q$ moment interacts with an electric field gradient $eq$ acting
at the site of the implanted nucleus in a single crystal stopper. The
$eqQ$ interaction causes the energy shift in the individual Zeeman
magnetic sublevels. Thus, the $Q$ moment is determined from the
measurement of the frequencies for the resonance transition between
the Zeeman $+$ quadrupole splitted sublevels, whose signal is detected
as a change in the $\beta$-ray asymmetry (the $\beta$-NQR method). The
$\beta$-rays emitted from the implanted nuclei were detected by
plastic-scintillator telescopes located above and below the stopper,
each consisting of three 1~mm-thick plastic scintillators. The up/down
ratio $R$ of the $\beta$-ray counts is written as
\begin{equation}
R  = a \frac{1 + v/c \cdot A_{\beta} P}{1 - v/c \cdot A_{\beta} P} 
\simeq a(1+2A_{\beta} P),
\label{EQ:UD}
\end{equation}
where $a$ is a constant factor representing asymmetries in the counter
solid angles and efficiencies, $v/c$ the velocity of the $\beta$
particle, $A_{\beta}$ the asymmetry parameter, and $P$ the 
$^{31}{\rm Al}$ nuclear spin polarization. Since we adopted a
high-energy portion of the $\beta$-ray energy spectrum 
($v/c \approx 1$) in the analysis, the $R$ ratio is approximated as in
the second expression in Eq.~(\ref{EQ:UD}). The adiabatic fast passage
(AFP) technique~\cite{ABRAGAM} was incorporated in order to pursue the
reversal, but not the destruction, of the spin polarization. By taking
a double ratio $R/R_{0}$ where $R_0$ is the value for $R$ measured
without the $B_1$ filed, the resonance frequency is derived from the
position of a peak or dip deviating from unity. An oscillating
magnetic field $B_{1}$ in a direction perpendicular to the external
field $B_{0}$ was applied with a pair of coils located outside a
vacuum jacket, in which the $\alpha{\mathchar`-}{\rm Al_{2}O_{3}}$
stopper was placed. In a first order perturbation theory, the
resonance frequency $\nu_{m, m+1}$ between magnetic sublevels $m$ and
$m+1$ of the nuclear spin $I$ under the combined Zeeman and quadrupole
interactions is given by  
\begin{eqnarray}
\nu_{m, m+1} = \nu_{\rm L} - \nu_{\rm Q}
  ( 3{\rm cos}^{2}\theta_{c\mbox{{\scriptsize -}}\mathrm{axis}} -1 )
  (2m+1)/4 \label{EQ:NuQ} \\
  \left( \nu_{\rm Q} = \frac{3}{2I(2I-1)} \cdot \frac{eqQ}{h} \right) 
  \nonumber
\end{eqnarray}
where $\nu_{\rm L}$ denotes the Larmor frequency, $eq$ the electric
field gradient along the {\it c}-axis (the additional term arising
from a deviation $\eta$ from the axial symmetry of the field gradient
tensor is omitted, since $\eta$ is reported to be
small~\cite{WO99_AL2O3}), 
$\theta_{c\mbox{{\scriptsize -}}\mathrm{axis}}$ the angle between the
{\it c}-axis and the $B_{0}$ field, and $eqQ / h$ the quadrupole
coupling constant. $Q$ and $h$ denote the $Q$ moment and the Planck's
constant, respectively. Inserting $I=5/2$ and
$\theta_{c\mbox{{\scriptsize -}}\mathrm{axis}}$ = 0 for the present
$^{31}{\rm Al}$ experiment, Eq.~(\ref{EQ:NuQ}) reads as
\begin{widetext}
\begin{eqnarray}
\nu_{m, m+1}(\nu_{\rm Q})
&=& \nu_{\rm L} - \frac{2m+1}{2} \nu_{\rm Q} \\
&=& \begin{cases}
\nu_{\rm L} + 2 \nu_{\rm Q} & \mbox{for}\ (m,m+1) = (-5/2, -3/2);
(\mbox{\ frequency\ ``$a$''}) \\
\nu_{\rm L} +   \nu_{\rm Q} & \mbox{for}\ (-3/2, -1/2)\mbox{, (``$b$'')} \\
\nu_{\rm L}          & \mbox{for}\ (-1/2, +1/2)\mbox{, (``$c$'')} \\
\nu_{\rm L} -   \nu_{\rm Q} & \mbox{for}\ (+1/2, +3/2)\mbox{, (``$d$'')} \\
\nu_{\rm L} - 2 \nu_{\rm Q} & \mbox{for}\ (+3/2, +5/2)\mbox{, (``$e$'')} \\
\end{cases}\label{EQ:NuQ_31Al}
\end{eqnarray}
where
\begin{eqnarray}
\nu_{\rm Q} = \frac{3}{20}\cdot\frac{eqQ}{h}. \label{EQ:NuQCoef_31Al}
\end{eqnarray}
\end{widetext}
The $B_1$ field was applied in $I$(2$I$+1) = 15 steps, each tuned to
one of the five transitions of Eq.~(\ref{EQ:NuQ_31Al}), in a sequence
{\it abcdeabcdabcaba} within the $B_1$ application period of 63~ms
duration. Actually, in the individual step the frequency $\nu$ of the
$B_1$ field was swept for the AFP method, over a frequency bin 
$\nu_{m,m+1}({\nu_{\rm Q}}^{\rm lower})$ $\rightarrow$
$\nu_{m,m+1}({\nu_{\rm Q}}^{\rm upper})$ that corresponded to a
$\nu_{\rm Q}$ region ${\nu_{\rm Q}}^{\rm lower}$ $\rightarrow$ 
${\nu_{\rm Q}}^{\rm upper}$. Details of the $B_1$ field sequence is
presented in Ref.~\cite{NAGAE_EMIS}. In evaluating
$\nu_{m,m+1}(\nu_{\rm Q})$ from Eq.~(\ref{EQ:NuQ_31Al}), we adopted a
value for the $^{31}{\rm Al}$ Larmor frequency, $\nu_{\rm L}$ =
5850(12)~kHz. This value was obtained from a $\beta$-NMR experiment on
$^{31}{\rm Al}$ in a Si crystal, carried out prior to the present
$Q$-moment measurement using the same apparatus and $B_0$ setting. 

\begin{figure}[tbhp]
\centering
\includegraphics[clip, width=0.4\textwidth]{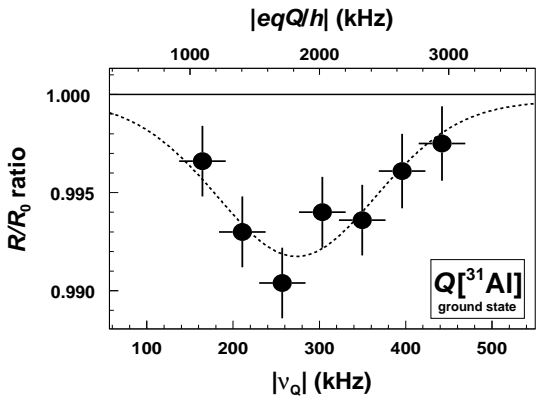}
\caption{An NQR spectrum obtained in an 
  $\alpha{\mathchar`-}{\rm Al_{2}O_{3}}$ crystal for the ground state
  of $^{31}{\rm Al}$. The $R/R_{0}$ ratio is plotted as a function of  
  $\nu_{\rm Q}$ or the corresponding quadrupole coupling constant
  $eqQ/h$. The vertical bar attached to the data point represents the
  statistical error due to $\beta$-counting statistics, while the
  horizontal bar indicates the width of $\nu_{\rm Q}$ frequency
  sweep. The result of the least-$\chi^2$ fitting analysis is shown by
  a dotted curve.} 
\label{FIG:NQR_31Al}
\end{figure}

Thus, the $\beta$-ray count ratio $R/R_{0}$ is expected to differ from
unity when the $B_1$ field sequence was executed for the $\nu_{\rm Q}$
region that includes the true $\nu_{\rm Q}$ value given by
Eq.~(\ref{EQ:NuQCoef_31Al}). Figure~\ref{FIG:NQR_31Al} shows the
measured $R/R_{0}$ ratio for $^{31}{\rm Al}$ in
$\alpha{\mathchar`-}{\rm Al_{2}O_{3}}$ as a function of $\nu_{\rm Q}$
(the $\beta$-NQR spectrum). The horizontal bar attached to the data
point (filled circle) indicates the $\nu_{\rm Q}$ region over which
the $B_1$ field frequency was swept. The vertical bar represents the
error in $R/R_{0}$ arising from the $\beta$-ray counting
statistics. The $R/R_{0}$ value at the dip bottom shows a displacement
of 5.3 standard deviation from unity, clearly indicating the
occurrence of the AFP spin reversal. The width of the dip, however,
seems to be substantially broader than that expected for a single
value of $eq$. 

The obtained NQR spectrum was fitted with a function
\begin{eqnarray}
F(\nu_{\rm Q}) = a \int
  G_{\sigma}(\xi) \cdot 
         {\cal F_{\rm AFP}}(\nu_{\rm Q} -{\nu_{\rm Q}}^{(0)}- \xi)
  {\rm d}\xi + b
  \label{EQ:FITTING}
\end{eqnarray}
with four free parameters ${\nu_{\rm Q}}^{(0)}$, $\sigma$, $a$, and
$b$ to be determined through the fitting. The $F(\nu_{\rm Q})$
function is a Gaussian convolution of a theoretical shape function
${\cal F_{\rm AFP}}(x)$ of a detuning $x$ (i.e., $f(x)$ of Eq.~(4) in
Ref.~\cite{OGAWA_Q17B}. The width of ${\cal F_{\rm AFP}}(x)$ was
evaluated to be 98~kHz (FWHM).) representing the expected shape of
$\beta$-NQR spectrum in the AFP mode for a single value of electric
field gradient $eq$. The parameter ${\nu_{\rm Q}}^{(0)}$ represents
the position of the dip, from which the quadrupole coupling constant
$eqQ/h$ will be deduced through Eq.~(\ref{EQ:NuQCoef_31Al}). The
parameter $\sigma$ is the width of the Gaussian function
$G_{\sigma}(\xi)$ $\equiv$ 
$(\sqrt{2\pi}\sigma)^{-1} \exp(-{\xi}^2 / 2\sigma^2)$, representing
extrabroadning effects that are not included in the function 
${\cal F_{\rm AFP}}(x)$. Extrabroadning may arise from distribution of
$eq$ value due to lattice defects or impurities in the stopper crystal
and misalignments in the $c$-axis orientation and $\nu_{\rm L}$
setting. In the present analysis such effects are expressed as a
finite value of $\sigma$. The extrabroadening was not included in the
preliminary reports~\cite{NAGAE_EMIS,UENO_Q31Al}.

From the fitting analysis of the NQR spectrum, we obtained 
${\nu_{\rm Q}}^{(0)}$ = 274(18) kHz for the dip position. The
resulting curve $F(\nu_{\rm Q})$ is shown in Fig.~\ref{FIG:NQR_31Al}
by a dotted line. Although the parameter ${\nu_{\rm Q}}^{(0)}$ is
determined with a rather small uncertainty 
($\delta^{\rm fit}{\nu_{\rm Q}}^{(0)}$ = 18~kHz) from the fitting
procedure, the actual spectrum is dominated by a much larger
broadening, $\sigma$ = 76~kHz, whose origins are not well
pinpointed. We therefore take into account the extrawidth $\sigma$ as
an independent error, and assign an experimental error 
$\delta {\nu_{\rm Q}}^{(0)}$ = 78~kHz, as shown in
Table~\ref{TBL:GOSA}. As a result, we obtain a quadrupole coupling
constant $| \nu_{\rm Q} |$ = $3/20\cdot|eqQ/h|$ = 274(78)~kHz, or
$|eqQ/h(^{31}{\rm Al})|$ = 1.83(52)~MHz. The $Q$ moment of 
$^{31}{\rm Al}$ is deduced from the relation $|Q(^{31}{\rm Al})|$ =
$|Q(^{27}{\rm Al}) \cdot 
\left( eqQ/h(^{31}{\rm Al}) \right) / 
\left( eqQ/h(^{27}{\rm Al}) \right) |$,
where $Q(^{27}{\rm Al})$ and $eqQ/h(^{27}{\rm Al})$ denote the {\it Q}
moment of $^{27}$Al and the quadrupole coupling constant of $^{27}$Al
in $\alpha{\mathchar`-}{\rm Al_{2}O_{3}}$, respectively. By inserting
the recently reported $Q$ moment $Q(^{27}{\rm Al})$ =
146.6(10)~$e$mb~\cite{KE99_Q27AL} and quadrupole coupling constant
$eqQ/h(^{27}{\rm Al})$ = 2389(2)~kHz~\cite{GR90_eqQ70K}, the
ground-state $Q$ moment $^{31}{\rm Al}$ is determined as 
$| Q_{\rm exp}(^{31}{\rm Al}) |$ = 112(32)~$e$mb.

\begin{table}[tbhp]
\caption{\label{TBL:GOSA}
Uncertainties taken into account for the determination of 
$| Q_{\rm exp}(^{31}{\rm Al}) |$ moment. Those uncertainties were
converted into the corresponding $\nu_{\rm Q}$ frequency.}
\begin{ruledtabular}
\catcode`?=\active \def?{\phantom{0}}
\begin{tabular}{lc}
Resonance $\nu_{\rm Q}$  & 274 (kHz) \\
\hline
( Statistical error )\\
\hspace*{7mm}Fitting error   & ?18 (kHz) \\
\vspace*{-1.5ex}
\\
( Systematic errors )\\
\hspace*{7mm}Ambiguity from the resonance width & ?76 (kHz) \\
\hspace*{7mm}Uncertainty of the electric field gradient & ??2 (kHz) \\
\hspace*{7mm}Ambiguity of the 
             $\theta_{c\mbox{{\scriptsize -}}\mathrm{axis}}$-angle 
             setting & 0.1 (kHz) \\
\hline
Total          & ?78 (kHz) \\
\hspace*{7mm} $\rightarrow$ $Q_{\rm exp}(^{31}{\rm Al})$  
              & 112 $\pm$ 32 ($e$mb)????? \\
\end{tabular}
\end{ruledtabular}
\end{table}

\begin{figure}[tbhp]
\centering
\includegraphics[clip, width=0.4\textwidth]{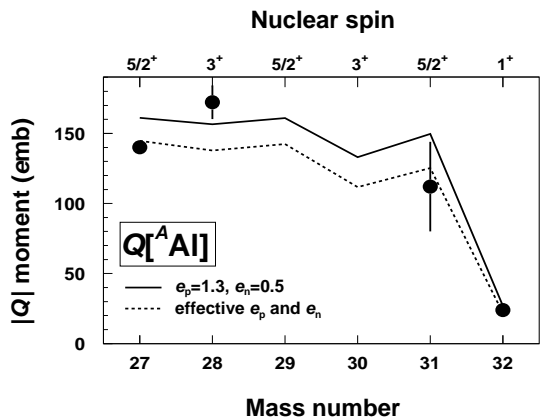}
\caption{The experimental (filled circle) and theoretical (solid and
  dotted lines) $Q$ moments of neutron-rich aluminum isotopes as a
  function of mass number, whose nuclear spins are also
  shown. Theoretical values are obtained from shell-model calculations
  within the {\it sd} shell with the USD interaction, using the
  constant effective charges $e_{\rm p}$ = 1.3 and $e_{\rm n}$ = 0.5
  (solid line) and isospin-dependent effective
  charges~\cite{OGAWA_Q17B} (dotted line).} 
\label{FIG:MassDep_QAl}
\end{figure}

In Fig.~\ref{FIG:MassDep_QAl}, the experimentally known $Q$ moments
for the neutron-rich aluminum isotopes including the present data are
plotted as a function of the mass number $A$. 
Also, the results of shell-model calculations within the {\it sd}
shell~\cite{W,OXBASH} are shown by the solid line.
The calculations reproduce the observed trend of the $Q$ moments in the
$^{27-32}$Al region fairly well:
$| Q_{\rm exp} |$ stays almost constant at $| Q_{\rm exp} |$  $\sim$
150~$e$mb, but suddenly decreases to a very small value at $A$ =
32~\cite{KAMEDA_Q32Al}. These calculations have employed effective
charges $e_{\rm p}$ = 1.3 for proton and $e_{\rm n}$ = 0.5 for
neutron. One could include the effect of isospin dependence of the
effective charges, which had been pointed out in
Ref.~\cite{BOHR_MOTTELSON} and was observed experimentally for the
first time in the $Q$ moments of boron isotopes~\cite{OGAWA_Q17B}. The
dotted line shows the calculated $Q$ with the effective charges varied
with isospin according to the expression given in
Ref.~\cite{OGAWA_Q17B}. The use of the isospin-dependent $e_{\rm p}$
and $e_{\rm n}$ reduces the calculated $Q$ by 10 $\sim$ 15~\% in the
$^{27-32}$Al isotopes, and improves the agreement particularly in the
$^{31}{\rm Al}$ $Q$ moment, although the experimental error in
$Q(^{31}{\rm Al})$ is not small. Finally, in contrast to the
approximate accordance of the {\it sd}-shell calculations with
experiment, an anticipation that the deformation might set in
somewhere along the chain of Al isotopes proves to be not the case at
least until $A$ = 32, since sizes of the experimental $Q$ presented in
Fig.~\ref{FIG:MassDep_QAl} are much smaller than those expected for
deformations of $\beta$ $\sim$ 0.5 occurring in the neighboring
nucleus $^{30}$Mg~\cite{CH01}.  


In summary, the ground-state  $Q$ moment of $^{31}$Al has been
determined by the $\beta$-NQR method, using the fragmentation-induced
spin polarization. The obtained $Q$ for $^{31}{\rm Al}$ as well as
known $|Q|$ values for other neutron-rich aluminum isotopes were found
to be well explained by shell-model calculations within the {\it sd}
shell. Viewing also that the magnetic moment of $^{31}$Al recently
determined~\cite{BO02} is explained with the same calculations, it is
concluded that $^{31}{\rm Al}$ is located outside the  
{\it island of inversion}.  

The authors wish to thank the staff of the RIKEN Ring Cyclotron for
their support during the experiment. They would like to thank
Dr.~E.~Yagi for useful help and advice with the X-ray diffraction
analysis of the $\alpha{\mathchar`-}{\rm Al_{2}O_{3}}$ sample. The
authors D.~N. and K.~S. are grateful for the Junior Research Associate
Program in RIKEN. This experiment was carried out at the RI Beam
Factory operated by RIKEN Nishina Center and CNS, University of Tokyo
under the Experimental Program No. R398n(5B).

\end{document}